\documentclass[preprint,showpacs,preprintnumbers,amsmath,amssymb]{revtex4}

\usepackage{graphicx}
\usepackage{dcolumn}
\usepackage{bm}

\renewcommand{\t}[1]{\tilde{#1}}
\newcommand{\an}[1]{\left\langle #1 \right\rangle}
\def\vec#1{\mbox{\boldmath $#1$}}
\newcommand{\Rcpd}{R_c(D)}

\newcommand{\vy}{ \vec{y} }
\newcommand{\vs}{ \vec{s} }

\begin{document}

\preprint{APS/123-QED}

\title{Statistical Mechanical Approach to Lossy Data Compression:
Theory and Practice}

\author{Tadaaki Hosaka}
 \email{hosaka-t@aist.go.jp}
\affiliation{%
National Institute of Advanced Industrial
Science and Technology\\
Tsukuba, Japan 3058568 
}%

\author{Yoshiyuki Kabashima}
\affiliation{
Department of Computational Intelligence and Systems Science\\
Tokyo Institute of Technology, Yokohama, Japan 2268502
}%

\date{\today}

\begin{abstract}
The encoder and decoder for lossy data 
compression of binary memoryless sources 
are developed on the basis of a 
specific-type nonmonotonic perceptron. 
Statistical mechanical analysis indicates that 
the potential ability of the perceptron-based 
code saturates the theoretically achievable limit in most 
cases although exactly performing the compression is 
computationally difficult. To resolve this difficulty, 
we provide a computationally tractable 
approximation algorithm using belief
propagation (BP), which is a current standard 
algorithm of probabilistic inference. Introducing several 
approximations and heuristics, the BP-based algorithm 
exhibits performance that is close to the achievable limit
in a practical time scale in optimal cases. 
\end{abstract}

\maketitle


\section{\label{sec:intro}Introduction}
Lossy data compression is a core technology of
contemporary broadband
communication.
The best achievable performance of lossy compression 
is theoretically provided by rate-distortion theorems,
which were first proved by Shannon for memoryless 
sources \cite{RD}. Unfortunately, Shannon's proofs 
are not constructive and suggest
few clues for how to design practical codes.
Consequently, no practical schemes saturating 
the potentially optimal performance of lossy 
compression represented by the 
rate-distortion function (RDF) have been found yet,
even for simple information sources. Therefore, the 
quest for better lossy compression 
codes remains a major problem in the field of information 
theory (IT).

Recent research on error correcting codes has 
revealed a similarity between 
IT and statistical mechanics (SM) 
of disordered systems~\cite{Nishimori}. 
Because it has been shown that methods from 
SM can be useful to analyse subjects of IT, 
it is natural to expect that a similar approach might 
also bring about novel developments in lossy compression.

This research is promoted by such a motivation.
Specifically, we propose a simple compression
code for uniformly biased binary data devised on 
input-output relations of a perceptron.
Theoretical evaluation based on the replica method (RM)
indicates that this code potentially saturates the RDF in most cases
although exactly performing the compression is 
computationally difficult. To resolve this difficulty, 
we develop a computationally tractable algorithm 
based on belief propagation (BP)~\cite{Pearl},
which offers performance that approaches the RDF in a practical 
time scale when optimally tuned.


\section{Lossy data compression}
We describe a general scenario for lossy data compression 
of memoryless sources.
Original data is denoted as 
$\vec{y} = (y^1,y^2,\ldots,y^M)$, 
which is
assumed to comprise 
a sequence of $M$ discrete or continuous random
variables that are generated independently 
from an identical stationary distribution $p(y)$.
The purpose of lossy compression is 
to compress $\vec{y}$ into a binary expression 
$\vec{s} = (s_1,s_2,\ldots,s_N)~(s_i \in \{+1,-1\})$, 
allowing a certain amount of {\em distortion} between the 
original data $\vec{y}$ and its representative 
vector $\t{\vec{y}} 
= (\t{y}^1,\t{y}^2,\ldots,\t{y}^M)$
when $\t{\vec{y}}$ is retrieved from $\vec{s}$.

In this study, distortion 
is measured using a distortion function that is assumed to be 
defined in a component-wise manner as 
$\mathcal{D}(\vec{y}, \t{\vec{y}}) = \sum_{\mu=1}^M
d(y^\mu, \t{y}^\mu)$, where $d(y^\mu, \t{y}^\mu) \ge 0$.
A code $\mathcal{C}$ is specified by a map
$\t{\vec{y}}(\vec{s}; \mathcal{C}):\vec{s} 
\to \t{\vec{y}}$, which is
used in the restoration phase.
This map also reasonably determines the compression phase as 
\begin{eqnarray}
\vec{s}(\vec{y}; \mathcal{C}) = 
\underset{\vec{s}}{\mathrm{argmin}}
\{ \mathcal{D}(\vec{y}, \t{\vec{y}}(\vec{s}; \mathcal{C}))\},
\label{eq:selecting_s}
\end{eqnarray}
where $\mathrm{argmin}_{\vec{s}} \{ \cdots\}$ represents 
the argument $\vec{s}$ that minimises $\cdots$. 
When ${\mathcal C}$ is generated from 
a certain code ensemble, typical codes 
satisfy the fidelity 
criterion 
\begin{eqnarray}
\frac{1}{M}
\mathop{\rm min}_{\vec{s}}
\{\mathcal{D}(\vec{y},\t{\vec{y}}(\vec{s}; \mathcal{C}) \} 
= \frac{1}{M}\mathop{\rm min}_{\vec{s}}
\left\{ \sum_{\mu=1}^M d(y^\mu, \t{y}^\mu(\vec{s};\mathcal{C}))
\right\} < D, 
\label{eq:fidelity}
\end{eqnarray}
for a given permissible distortion $D$ and typical original data 
$\vec{y}$ with probability $1$
in the limit $M,N \to \infty$ maintaining 
the coding rate $R \equiv N/M$ constant, if and only if $R$
is larger than a certain critical rate $\Rcpd$ that is termed 
the \textit{rate-distortion function}.

However, for finite $M$ and $N$,
any code has a finite probability 
$P_{\rm \tiny F}$ of breaking the 
fidelity (\ref{eq:fidelity}),
even for $R>\Rcpd$. 
Similarly, for $R <\Rcpd$, 
Eq. (\ref{eq:fidelity}) 
is satisfied with a certain probability $P_{\rm \tiny S}$. 
For reasonable code ensembles, 
the averages of these probabilities
are expected to decay exponentially with respect to $M$ 
when the data length $M$ is sufficiently large. 
Therefore, the two \textit{error exponents}
$\alpha_A (D,R) = \lim_{M \to \infty} 
- (1/M)
\ln \left \langle P_{\rm \tiny F} 
\right \rangle_{\cal C}$ for $R >\Rcpd$ 
and $\alpha_B (D,R) = \lim_{M \to \infty} 
-(1/M) \ln\left \langle  P_{\rm \tiny S} 
\right \rangle_{\cal C}$ for $R <\Rcpd$,
where $\langle \cdots \rangle_\mathcal{C}$ represents the average over
the code ensemble,
can be used to characterise
the potential ability of the ensemble of finite 
data lengths.


\section{Compression by perceptron and theoretical evaluation}

It is conjectured that 
the components of $\vec{s}$
are preferably unbiased and uncorrelated
in order to minimise loss of information in the original 
data from a binary information source, which implies 
that the entropy per bit in $\vec{s}$ must be maximised. 
On the other hand, 
in order to reduce the distortion, the representative
vector $\tilde{\vec{y}}(\vec{s}; \mathcal{C})$ should be placed close to 
the typical sequences of the original data that are biased. 
Unfortunately, it is difficult to construct a code that satisfies 
these two requirements using only linear transformations 
over the Boolean field because a linear transformation 
generally reduces statistical bias in sequences, which implies that 
one cannot produce a biased representative vector 
from the unbiased compressed sequence.

One method to design a code that has the above properties 
is to introduce a nonlinear transformation. 
A perceptron provides a simple 
scheme for carrying out this task. 
To specify lossy data compression codes for binary 
original data $\vec{y} \in \{ +1, -1\}^M$ generated from a memoryless
source, 
we define a map by utilising perceptrons from 
the compressed expression 
$\vec{s} \in \{+1,-1\}^N$ to
the representative sequence 
$\tilde{\vec{y}}(\vec{s}; \mathcal{C}) \in \{+1, -1\}^M$ as
\begin{eqnarray}
\tilde{y}^\mu(\vec{s}; \{\vec{x}^\mu \})=f \left(\frac{1}{\sqrt{N}} 
\sum_{i=1}^N x_i^\mu s_i \right),~~~~~(\mu=1,2,\ldots,M)
\label{perceptron}
\end{eqnarray}
where $f(\cdot)$ is a function for which 
the output is limited to $\{+1, -1\}$
and $\vec{x}^{\mu=1,2,\ldots,M}$ are 
randomly 
predetermined $N$-dimensional vectors 
that are generated from 
an $N$-dimensional 
normal distribution 
$P(\vec{x}) = \left (\sqrt{2 \pi} \right )^{-N} \exp 
\left [-|\vec{x}|^2/2 \right ]$. 
These vectors are known to the encoder and 
decoder.
We adopt an output function $f_k(u) = 1$
for $|u|<k$, and $-1$ otherwise,
which eventually offers optimal performance.

We measure the distortion by the Hamming distance 
$\mathcal{D}( \vec{y}, \t{\vec{y}}(\vec{s}; \{ \vec{x}^\mu \})) = 
\sum_{\mu=1}^M \left[  \left\{
1 - y^\mu \cdot f_k \left( 
\sum_{i=1}^N x_i^\mu s_i / \sqrt{N} \right) \right\} / 2  \right].
$
Then, the compression phase for the given data $\vec{y}$ can be defined as 
finding a vector $\vec{s}$ that minimises the resulting distortion 
$\mathcal{D}(\vec{y},\tilde{\vec{y}}(\vec{s}; \{ \vec{x}^\mu \} ))$, 
and the retrieval process can be performed easily using
Eq. (\ref{perceptron}) from a given sequence $\vec{s}$.
The performance evaluation 
has been investigated theoretically from the
perspective of SM, 
which is not specialised for this
perceptron-based code. Rather, it is a general one as mentioned briefly below.

Let us regard the distortion function 
$\mathcal{D}(\vec{y}, \t{\vec{y}} (\vec{s}; \mathcal{C}))$ 
as the Hamiltonian for the dynamical variable $\vec{s}$, 
which also depends on predetermined
variables $\vec{y}$ and $\mathcal{C}$. 
The resulting distortion (per bit) for a given $\vec{y}$ and
$\mathcal{C}$
is represented as
$\lambda(\vec{y},\mathcal{C}) = \mathrm{min}_{\vec{s}}
\{ M^{-1} \mathcal{D}(\vec{y}, \t{\vec{y}}(\vec{s}; \mathcal{C}))\}$.
We start with a statistical mechanical inequality
\begin{eqnarray}
e^{-M \beta \lambda(\vec{y}, \mathcal{C})} \le 
\sum_{\vec{s}}e^{-\beta \mathcal{D}(\vec{y}, \t{\vec{y}} 
(\vec{s}; \mathcal{C}))}
=Z(\beta;\vec{y},\mathcal{C})=e^{-M\beta f(\beta;\vec{y},\mathcal{C})}, 
\label{ineq}
\end{eqnarray}
which holds for any sets of $\beta \!>\! 0, \vec{y}$ and $\mathcal{C}$. 
The physical implication of this is 
that the ground state energy $\lambda(\vec{y}, \mathcal{C})$ (per component) 
is lower bounded by the free energy 
$f(\beta; \vec{y}, \mathcal{C})$ (per component)
for an arbitrary temperature $\beta^{-1} >0$. 
In particular, the free energy $f(\beta; \vec{y}, \mathcal{C})$ agrees with 
$\lambda(\vec{y}, \mathcal{C})$ 
in the zero temperature limit $ \beta \to \infty$, which is the key for
the analysis.

The distribution of the free energy $P(f; \beta)$
is expected to peak at its typical value of 
\begin{eqnarray}
f_t(\beta) = - \frac{1}{M \beta}
\an{ \ln Z(\beta; \vec{y}, \mathcal{C}) }_{\vec{y}, \mathcal{C}}
= \lim_{n \to 0} \frac{-1}{M \beta} \frac{\partial}{\partial n}
\ln \an{ Z^n(\beta; \vec{y}, \mathcal{C})}_{\vec{y}, \mathcal{C}},
\end{eqnarray}
where $\left \langle \cdots \right \rangle_{\vec{y}, 
\mathcal{C}}$ denotes
the average over $\vec{y}$ and $\mathcal{C}$,
and decays exponentially away from $f_t(\beta)$ as 
$P(f;\beta) \sim \exp[ - M c(f, \beta)]$ 
for large $M$. 
Here, we assume that $c(f, \beta)\ge 0$ is 
a convex downward function that is minimised to $0$ at $f=f_t(\beta)$. 
This formulation implies that, for $\forall{n} \in {\bf R}$, 
the logarithm of the moment of the partition function $Z(\beta;\vec{y},
\mathcal{C})$, 
$g(n, \beta) \equiv -M^{-1} \ln 
\an{Z^n(\beta; \vec{y}, \mathcal{C})}_{\vec{y}, \mathcal{C}}$, 
can be evaluated by the saddle point method as
\begin{eqnarray}
g(n, \beta) 
= \underset{f}{\mathrm{min} } \{ n \beta f + c(f, \beta) \}.
\label{eq:g_Legendre}
\end{eqnarray}
Based on the Legendre transformation (\ref{eq:g_Legendre}), 
$c(f, \beta)$ is assessed by the inverse 
transformation 
\begin{eqnarray}
c(f, \beta) = \underset{n}{ \mathrm{max} } 
\{ - n \beta f + g(n, \beta) \},
\label{eq:Legendre}
\end{eqnarray}
from $g(n, \beta)$, which can be evaluated using the RM
analytically extending expressions obtained 
for $n \in {\bf N}$ to $n \in {\bf R}$.

The above argument indicates that 
the typical value of the distortion 
averaged over the generation with respect to 
$\vec{y}$ and $\mathcal{C}$ 
can be evaluated as
\begin{eqnarray}
\an{ \lambda(\vec{y}, \mathcal{C}) }_{\vec{y}, \mathcal{C} } 
= \lim_{\beta \to \infty} \lim_{n \to 0} 
\frac{1}{\beta}\frac{\partial g(n,\beta)}{\partial n},
\label{typical_case}
\end{eqnarray}
and that 
the average error exponent 
$\alpha_{\{ A,B \}}(D,R)$, which is an abbreviation denoting
$\alpha_{A}(D,R)$ and $\alpha_{B}(D,R)$, can be assessed as
\begin{eqnarray}
\alpha_{ \{ A,B \} }(D,R) 
=
\lim_{\beta \to \infty} c(f\!=\!D, \beta)
=
\lim_{\beta \to \infty}
\left \{
-n \frac{\partial g(n, \beta)}{\partial n} + g(n, \beta)
\right \},
\label{exact_exponent}
\end{eqnarray}
where $n$
is a function of $\beta$ that is determined by the extremum condition
of Eq. (\ref{eq:Legendre}) as $\beta^{-1} \partial g(n,\beta) / 
\partial n = D$. 
Equations(\ref{typical_case}) and (\ref{exact_exponent}) 
constitute the basis of our approach.

When the above general framework was applied to the random code
ensemble, which is not a practical coding scheme, but can exhibit 
optimal performance, 
the theoretical limitations -- 
the RDF and optimal error exponents derived 
in IT \cite{Marton, Csiszar} --
were reproduced correctly  \cite{SPDSA, JPSJ}.
These results support the validity of our theoretical framework.
In addition to consistency with the existing 
results, 
we demonstrated the wide applicability of our framework 
for the perceptron-based code in \cite{JPSJ, PRE}, 
which indicated that the 
perceptron-based code can also saturate
the theoretical limitations in most cases.

\section{\label{sec:app}Algorithm based on belief propagation}

Calculation using the RM 
implies that the perceptron-based code potentially provides optimal
performance for binary memoryless sources.
However, this is insufficient when it is necessary to obtain
a compressed sequence $\vec{s}$ 
for a given finite length of original data $\vec{y}$.

For the perceptron-based code, the compression phase 
to follow the prescription (\ref{eq:selecting_s}) is computationally difficult
because it requires a comparison over $O(2^N)$ patterns
to extract the ground state for the relevant
Boltzmann distribution
$P_{\mathrm{B}}(\vec{s}|\vec{y}, \{\vec{x}^\mu \};\beta)
= 
\exp\left[ -\beta \mathcal{D}(\vec{y}, \t{\vec{y}}(\vec{s}; \{ \vec{x}^\mu \})
	     )\right] / Z(\beta; \vec{y}, \{ \vec{x}^\mu \}).$
The Boltzmann factor 
is rewritten here for the sake of subsequent
expressions as
\begin{eqnarray}
\exp
\left[ -\beta \mathcal{D}(\vec{y}, \t{\vec{y}}(\vec{s}; \{ \vec{x}^\mu
\}))\right] = \prod_{\mu =1}^M \Xi_{k, y^\mu}
\left(\frac{1}{\sqrt{N}} \sum_{i=1}^N x_i^\mu s_i \right),
\label{factorize}
\end{eqnarray}
where we define $\Xi_{k, y^\mu} (z) = \exp[- (\beta /2) \{
1- y^\mu \cdot f_k \left( z \right) \} ]$.
We require computationally tractable algorithms
that generate a probable state $\vec{s}$ 
from the Boltzmann distribution.

The BP is known as a promising approach for such tasks.
It is an iterative algorithm that efficiently 
calculates the marginal posterior probabilities
$P_{s_l}(s_l| \vec{y}, \{
\vec{x}^\mu \}; \beta) = \sum_{s_{i \ne l} }
P_{\mathrm{B}}(\vs| \vy, \{ \vec{x}^\mu \}; \beta )$ 
based on the property that the potential function is 
factored as shown in Eq. (\ref{factorize}). 
In general, the fixed point of this algorithm generally 
provides the solution of the 
Bethe approximation \cite{Kaba} known in SM.

\begin{figure}
  \begin{center}
  \includegraphics[width=8cm]{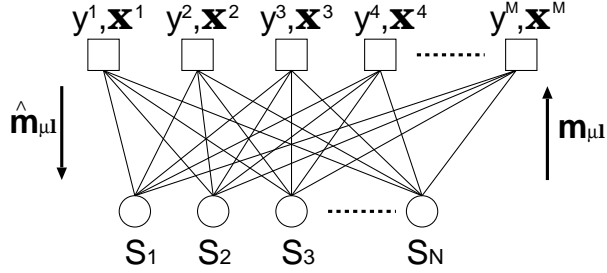}
  \end{center}
  \caption{Graphical representation of the dependency of variables for
 the perceptron-based code. Each
 pair of predetermined variables $(y^\mu, \vec{x}^\mu)$ is related
 with every bit of the compressed sequence.
           }
  \label{fig:bipartite}
\end{figure}

To introduce this algorithm to the current system, 
let us graphically 
describe this factorization, denoting
the predetermined variables ($y^\mu$ and $\vec{x}^\mu$) and
compressed sequence ($s_i$) by two kinds of nodes, then connecting them by
an edge when they are included in a common factor, which can be
expressed as a complete bipartite graph shown in FIG. \ref{fig:bipartite}.

On that graph, BP can be 
represented as an algorithm that passes messages between the
two kinds of nodes through edges as
\begin{eqnarray}
\hat{\rho}^{t+1}_{\mu l}(s_l) &=& \sum_{s_{i \ne l} } 
\Xi_{k, y^\mu}
\left(\frac{1}{\sqrt{N}} \sum_{i=1}^N x_i^\mu s_i \right)
          \prod_{j=1 (j \ne l)}^N \rho^t_{\mu j}(s_j),
\label{message2} \\
\rho^t_{\mu l}(s_l) &=& 
\prod_{\nu =1 (\nu \ne \mu)}^M \hat{\rho}^t_{\nu l}(s_l),
\label{message1} 
\end{eqnarray}
where $t=1,2,\ldots$ is an index for counting the number of updates.
The marginalised posterior at the $t$th update is given as
$P_{s_l}^t(s_l|\vec{y}, \{\vec{x}^\mu\}; \beta) \sim
\prod_{\mu=1}^M \hat{\rho}^t_{\mu
l}(s_l)$. 
Because $s_l$ is a binary variable, one can parameterize the above functions
as distributions
$
\hat{\rho}^t_{\mu l}(s_l) \sim
(1 + \hat{m}^t_{\mu l} s_l ) / 2,
\rho^t_{\mu l}(s_l) \sim 
(1+m^t_{\mu l} s_l) / 2,
$
and
$P^t_{s_l}(s_l|\vec{y}, \{\vec{x}^\mu\}; \beta)  
= (1 + m_l^t s_l) /2$.

Since the computational 
cost for the summation in Eq. (\ref{message2}) 
grows exponentially in $N$, 
it is extremely difficult to perform this algorithm exactly. 
However, because $\vec{x}^\mu$ are generated independently 
from $P(\vec{x}) = \left (\sqrt{2 \pi} 
\right )^{-N} \exp 
\left [-|\vec{x}|^2/2 \right ]$, 
this summation can be well approximated 
by a one-dimensional integral of 
a Gaussian distribution 
\cite{CDMA}, the centre
 and the variance of which are 
$\Delta_{\mu l}^t \equiv \left( \sum_{i \ne l} x_i^\mu m^t_{\mu i}
 \right) / \sqrt{N}$,
$1 - q_{\mu l}^t,$ where $q^t_{\mu l} \equiv 
\left\{ \sum_{i \ne l} \left( m^t_{\mu i}  \right)^2 \right\} / N$, 
respectively.
This approximation makes it possible to
carry out the belief updates
 (\ref{message2}) and (\ref{message1}) in a practical time scale,
 providing a set of self-consistent equations 
\begin{eqnarray}
\hat{m}^{t+1}_{\mu l } &=& 
\frac{x_l^\mu}{\sqrt{N}} \frac{\int Dz~ \Xi_{k, y^\mu}^\prime
\left( \Delta^t_{\mu l} + \sqrt{1 - q^t_{\mu l} } z \right) }
{
\int Dz~ \Xi_{k, y^\mu} 
\left( \Delta^t_{\mu l} + \sqrt{1-q^t_{\mu l}} z \right) 
},
\label{mhat} \\
m^t_{\mu l} &=& 
\tanh \left( \sum_{\nu = 1 (\nu \ne \mu) }^M 
\tanh^{-1} \hat{m}^t_{\nu l} \right),
\label{mmu}
\end{eqnarray}
where we define $Dz \equiv (dz / \sqrt{2 \pi}) \exp(-z^2 /2)$, and 
$f^\prime (x) \equiv d f(x) / dx$
for any function $f(x)$.
Employing these variables, the approximated posterior average of $s_i$
at the $t$th update can be computed as $m_l^t = \tanh \left( \sum_{\mu =1}^M
\tanh^{-1} \hat{m}_{\mu l }^t \right) 
\approx \tanh \left( \sum_{\mu=1}^M \hat{m}_{\mu
l }^t \right)$.


The number of variables can be further reduced to $O(M)$ when $M, N$ is
large by employing Eq. (\ref{mmu}) \cite{CDMA}.
One can approximately transform Eq. (\ref{mmu}) into 
$
m_{\mu l}^t 
\approx m^t_l - \left\{  1 - \left( m^t_l \right)^2 \right\} 
\hat{m}^t_{\mu l}.
$
Utilising this equation and taking into consideration that the influence of
each element in the sum is sufficiently small compared to the remains, 
the following approximations hold.
\begin{eqnarray}
\Delta^t_{\mu l} 
& \approx &
\frac{1}{\sqrt{N}} \sum_{i=1}^N x_i^\mu m^t_i 
- \frac{1}{\sqrt{N}} \sum_{i =1}^N  x_i^\mu 
\left\{ 1 - \left( m^t_i \right)^2 \right\} 
\hat{m}^t_{\mu i}
- \frac{1}{\sqrt{N}} x_l^\mu m^t_l,
\label{new_delta} \\
q^t_{\mu l} & \approx &
\frac{1}{N} \sum_{i=1}^N \left( m^t_i \right)^2 \equiv q^t.
\label{bp_q}
\end{eqnarray}
Because the last term in Eq. (\ref{new_delta})
is infinitesimal, the Taylor expansion is applicable  to 
Eq. (\ref{mhat}), providing
\begin{eqnarray}
\hat{m}^{t+1}_{\mu l} 
& \approx &
\frac{x^\mu_l}{\sqrt{N}}
\frac{\int Dz~ \Xi_{k, y^\mu}^\prime
\left( T_\mu^t \right) 
}
{
\int Dz~ \Xi_{k, y^\mu} 
\left( T_\mu^t \right) 
} 
-
\left\{
\frac{\int Dz~ \Xi_{k, y^\mu}^{\prime \prime}  
\left( T_\mu^t \right) 
}
{\int Dz~\Xi_{k, y^\mu}
\left( T_\mu^t \right) 
}
-
\left(
\frac{\int Dz~ \Xi_{k, y^\mu}^\prime
\left( T_\mu^t \right) 
}
{\int Dz~\Xi_{k, y^\mu}
\left( T_\mu^t \right) 
}
\right)^2
\right\}
\frac{ \left( x_l^\mu \right)^2  m^t_l }{N}, 
\label{new_mmul} \\
&& \left(T_\mu^t  \equiv 
   \Delta^t_{\mu} 
   - \frac{1}{\sqrt{N}} \sum_{i =1}^N  x_i^\mu \hat{m}^t_{\mu i} 
   \left\{ 1 - \left( m^t_i \right)^2 \right\} 
   + \sqrt{1-q^t } z \right), \notag  
\end{eqnarray}
where we define
$\Delta^t_\mu \equiv \left( \sum_{i=1}^N x^\mu_i m_i^t \right) / \sqrt{N}$.
It is important to notice that the second term of the right-hand side 
in Eq. (\ref{new_mmul}),
which 
can be negligible compared to the first term, 
becomes an influential element when the posterior average $m_l^t$ 
is calculated.

Using the new notations $a_\mu^t$ and $G^t$, 
the compression algorithm is finally expressed as 
\begin{eqnarray}
a^{t+1}_{\mu} 
&=& 
\frac{\int Dz~ \Xi_{k, y^\mu}^\prime
\left( U_\mu^t \right) }
{
\int Dz~ \Xi_{k, y^\mu} 
\left( U_\mu^t \right) 
}, 
\label{final_mhat} \\
G^t & = &
\sum_{\mu =1}^M
\left[ 
\frac{\int Dz~ \Xi_{k, y^\mu}^{\prime \prime}
     \left( U_\mu^t \right)
}
{\int Dz~\Xi_{k, y^\mu}
     \left( U_\mu^t \right)
} 
-
\left\{
\frac{\int Dz~ \Xi_{k, y^\mu}^\prime
     \left( U_\mu^t \right)
}
{\int Dz~\Xi_{k, y^\mu}
     \left( U_\mu^t \right)
} 
\right\}^2 \right],
\label{g_mu} \\
&& \left( U_\mu^t  \equiv 
\Delta^t_{\mu} - (1-q^t) a^t_\mu + \sqrt{1-q^t} z \right), \notag \\
m^t_l &=& \tanh \left[
\sum_{\mu=1}^M 
\frac{x_l^\mu}{\sqrt{N}} 
a^t_\mu 
- \frac{ G^{t-1} }{N} m_l^{t-1} 
\right].
\label{final_ml}
\end{eqnarray}
For the perceptron-based code, we can calculate
\begin{eqnarray}
\int Dz~\Xi_{k, y^\mu} (U_\mu^t)  &=& 
e^{-\beta} + (1-e^{-\beta}) \left\{ y^\mu H(w^t_{\mu -}) - y^\mu
			     H(w^t_{\mu +})
- ( y^\mu - 1 ) / 2 \right\},
\label{tau0} \\
\int Dz~\Xi^\prime_{k, y^\mu} (U_\mu^t)  &=& 
\frac{(1-e^{-\beta}) y^\mu }
{\sqrt{ 2 \pi ( 1-q^t ) } }
\left[
\exp \left\{ - \frac{ \left( w^t_{\mu -} \right)^2 } {2} \right\} 
-
\exp \left\{ - \frac{ \left( w^t_{\mu +} \right)^2 } {2} \right\} 
\right],
\label{tau1}\\
\int Dz~\Xi^{\prime \prime}_{k, y^\mu} (U_\mu^t)  &=& 
\frac{(1- e^{-\beta}) y^\mu }{\sqrt{2\pi}(1-q^t)}
\left[
w^t_{\mu -}
\exp \left\{ - \frac{ \left( w^t_{\mu -} \right)^2 }{2} \right\}
- 
w^t_{\mu +}
\exp \left\{ - \frac{ \left( w^t_{\mu +} \right)^2 }{2} \right\}
\right],
\label{tau2}
\end{eqnarray}
where
\begin{eqnarray}
w_{\mu \pm}^{t} = \frac{\pm k - \Delta_\mu^t + ( 1-q^t) a_{\mu}^t }
{\sqrt{1-q^t}},~~~~~
H(x) = \int_x^\infty \frac{dz}{\sqrt{2 \pi}} \exp 
\left(- \frac{z^2}{2} \right).
\notag
\end{eqnarray}

The exact solution of $m^t_l~(\forall l)$ 
trivially vanishes because of the mirror symmetry $\Xi_{k, y^\mu}(U_\mu^t)=
\Xi_{k, y^\mu}(-U_\mu^t)$. 
This fact implies that one cannot 
determine the more probable 
sign of $s_l$ 
even if the update iteration is successful. 
A similar phenomenon was also reported in 
codes of another type~\cite{Murayama, Mezard_code}. 
To resolve this problem, we heuristically 
introduce an inertia term that has been employed 
for lossy compression of an unbiased source \cite{Murayama}, 
in Eq. (\ref{final_ml}) as $m^t_l = \tanh \left[
\sum_{\mu=1}^M
x_l^\mu a_\mu^t / \sqrt{N} 
- G^{t-1} m^{t-1}_l / N
+ \tanh^{-1} \left( \gamma m^{t-1}_l \right)
\right],$
where $\gamma$ is a constant ($0 \le \gamma \le 1$).

\begin{figure}
  \begin{center}
  \includegraphics[width=9cm]{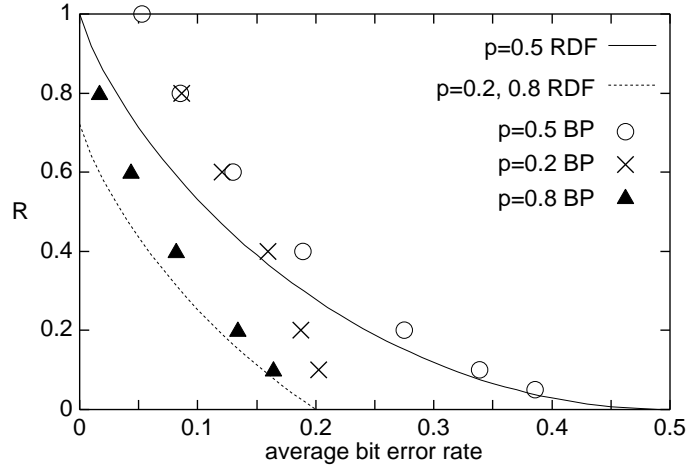}
  \end{center}
  \caption{Compression performance of BP for $p=0.2, 0.5$ and $0.8$.
    In the region of the low compression rate, the performance
 approaches the RDF in the case of $p=0.5$ and $0.8$.
           }
  \label{fig:bp}
\end{figure}

Experimental results are shown in FIG. \ref{fig:bp}
together with the RDFs.
Given original data generated from
a binary stationary distribution $P(y^\mu\!=\!+1) = 1 - P(y^\mu\!=\!-1) = p$
and vector $\{ \vec{x}^\mu \}$,
we compressed the original data into a shorter sequence using the
BP.
The final value of $s_l$ was determined as $s_l = \mathrm{sgn}[m_l^t]$.
The values of $k$ and $\beta$ were set to theoretically optimal 
values evaluated from the RM-based analysis; the value of $\gamma$ was
determined by trial and error.
In the figure, the bit error rates averaged over 100 runs are plotted
as a function of the compression rate $R$ for bias 
$p=0.2, 0.5$ and $0.8$.
While the compressed sequence was fixed to $N=1000$ bits ($N=500$ when $R \le
0.2$), the length of the original data was adjusted in accordance with the
compression rate.
We stopped the iteration at the 35th update and determined the
compressed sequence from the result at that time, even if the algorithm did not
converge.
As the compression rate becomes smaller,
the performance approaches the RDF in the case of
$p=0.5$ and $0.8$.
In particular, the performance for $p=0.5, R \le 0.4$ is superior to results
reported in the IT literature as a binary memoryless source
\cite{Yang}.
However, 
the results for $p=0.2$ yield poor performance 
compared to those for $p=0.8$, even though
the situation from the perspective of information 
is the same as $p=0.8$, which might be the result of asymmetric 
influences of input-output relations between those two cases.
Improvement of this behaviour is a subject of future work.

\section{\label{sec:summary}Summary}
We have investigated the performance of lossy data compression 
for uniformly biased binary data.
Analyses based on the RM 
indicate the great potential of the
perceptron-based code, which is also partially confirmed by a
practically tractable algorithm based on BP.
A close relationship between the macroscopic dynamics 
of BP and the replica analysis, was recently 
reported \cite{CDMA}. Investigation in such 
a direction in the current case is under way.

\begin{acknowledgements}
This research was partially supported by 
Grants-in-Aid Nos. 164453 (TH), 14084206 and 17340116 (YK)
from JSPS/MEXT, Japan. 
\end{acknowledgements}


\end{document}